\documentclass[onecolumn,preprintnumbers,footinbib,tightenlines,superscriptaddress]{revtex4}

\pdfoutput=1
\usepackage[english]{babel}
\usepackage{amsmath,amssymb,amsfonts,bm,bbm,slashed,subdepth}
\usepackage{graphicx}
\usepackage[sort&compress]{natbib}
\usepackage{xcolor}
\usepackage[normalem]{ulem}
\usepackage{hyperref}
\usepackage{cleveref}
\definecolor{red}{rgb}{1.0, 0, 0}
\usepackage{enumerate}
\usepackage{epsfig, subfigure}
\usepackage{setspace}
\usepackage{booktabs, tabularx}
\usepackage{units}
\usepackage{placeins}
\usepackage{multirow}
\usepackage{mathtools}
\usepackage{float}

\renewcommand{\thesection}{\arabic{section}}
\renewcommand{\thesubsection}{\thesection.\arabic{subsection}}
\renewcommand{\thesubsubsection}{\thesubsection.\arabic{subsubsection}}

\makeatletter
\renewcommand{\p@subsection}{}
\renewcommand{\p@subsubsection}{}
\makeatother

\newcommand{\sqq}{s_{q\overline{q}} }
\newcommand{\sgg}{s_{\gamma\gamma} }

\begin{document}

\preprint{MDPI Proceedings 2019, 13(1), 1; https://doi.org/10.3390/proceedings2019013001}
\title{Effective Field Theory Treatment of Monopole Production by Drell--Yan and Photon Fusion for Various Spins $^{\dagger}$}
\author{Stephanie Baines}
\email{stephanie.baines@kcl.ac.uk}
\affiliation{Theoretical Particle Physics and Cosmology Group, Department of Physics,King's College London, Strand, London WC2R 2LS, UK;
}

\begin{abstract} 
A resolution over the existence of magnetic charges has eluded the high energy physics community for centuries, and their search has gained momentum as recent models predict these  may be observable at current colliders. They appear in field theories in two forms: the widely studied but heavily suppressed monopole with structure (soliton) and the not-so-well-covered point-like monopole. The latter was first proposed by Dirac as the source of a singular magnetic field and in effect symmetrises Maxwell's equations. Following this line of research, work by S. Baines et al. analysed these sources as matter fields that carry spins 0, $\frac{1}{2}$, or 1, in an effective field theory that is perturbative for monopoles produced at threshold where the coupling strength $g(\beta)$ is suppressed. All three cases are currently under investigation by the MoEDAL collaboration at CERN, and the theoretical expressions for kinematic distributions proposed in this work serve as guides to these searches. 
The cross section distributions in each case are derived from a \emph{U}(1) invariant gauge theory. It is not assumed that, like the electron, the monopole's magnetic moment is generated through spin interactions at minimal coupling, as it may be quite large. Instead, the analytical expressions in the spin $\frac{1}{2}$ and $1$ cases are kept completely general through the inclusion of a phenomenological parameter $\kappa$, related to the gyromagnetic ratio $g_R=1+\kappa$. In fact, the inclusion of this parameter gives the effective theory validity in the high energy limit if the magnetic coupling scales with the particle's velocity $\beta=\frac{v}{c}$.
\end{abstract}

\maketitle

{\small \setlength{\parindent}{0em}Presented at the 7th International Conference on New Frontiers in Physics (ICNFP 2018), Crete, Greece, 4--12~July~2018.\par\setlength{\parindent}{0em}\vspace{5pt} \emph{Published:} 14 May 2019}\\

{\bf Introduction}\\
The theory of electromagnetism as formulated by Maxwell in 1873 is one of the most successful theories of nature, surviving tests of general relativity and quantum mechanics. While it is accepted that the model naturally incorporates electric charges, an isolated magnetic charge remains a concept useful only for mathematical convenience, without a physical interpretation \cite{kMilton}. 
Monopole physics has been a source of controversy since its formal conception by Dirac in 1931 \cite{Dirac1,Dirac2}.  Dirac proposed that a single valued quantum mechanical wavefunction with a singular phase functional would manifest its singularity as the presence of a magnetic source. The singularity corresponded to a string whose orientation represented a gauge choice. Despite attempts by Weinberg, Schwinger, Zwanziger, and others \cite{Weinberg, Schwinger, Zwanziger}, observables derived from this model of point-like monopoles remained both gauge-dependent and Lorentz-violating.

There has also been much success in deriving a topological structure with a net magnetic charge in gauge theories of scalar fields with spontaneous symmetry breaking. The first was the t'Hooft--Polyakov monopole, derived from a broken $SU(2)$ gauge theory in the adjoint representation. This was recently followed by the discovery of the non-trivial second homotopy of the Standard Model, which originates from a residual $CP^1$ symmetry, by Cho and Maison \cite{ChoMaison}, although it lacked a finite solution. This divergence was resolved by extending the Standard Model using a string-inspired Born--Infeld action in the hypercharge sector \cite{ArunKob}. The Dirac string was interpreted as the axis along which the $U(1)$ electromagnetic potential was singular. That said, this monopole solution was derived from a Lorentz-invariant theory and so the soliton must also be. Otherwise, this would signal a fundamental breakdown in the analytical techniques of  quantum theory. Hence, as these solitonic monopoles are extended to objects that recover a point-like interpretation at distances far from the structure's core,  there appeared to be a paradox.
The questions of Lorentz and gauge invariance were recently resolved by a re-summation of soft emissions in scattering processes \cite{Terning} in a toy model of monopoles involving perturbatively small magnetic couplings. But already in 1978, Urrutia showed that monopole-charge-particle scattering in a limited region of phase space was gauge invariant in the zeroth-order eikonal approximation \cite{Urrutia}. In \cite{fullreport}, it is thus assumed that the effective $U(1)$ theories for monopoles emerge from such gauge- and Lorentz-invariant considerations. 

\begin{figure}[H]
\centering
\includegraphics[scale=0.5]{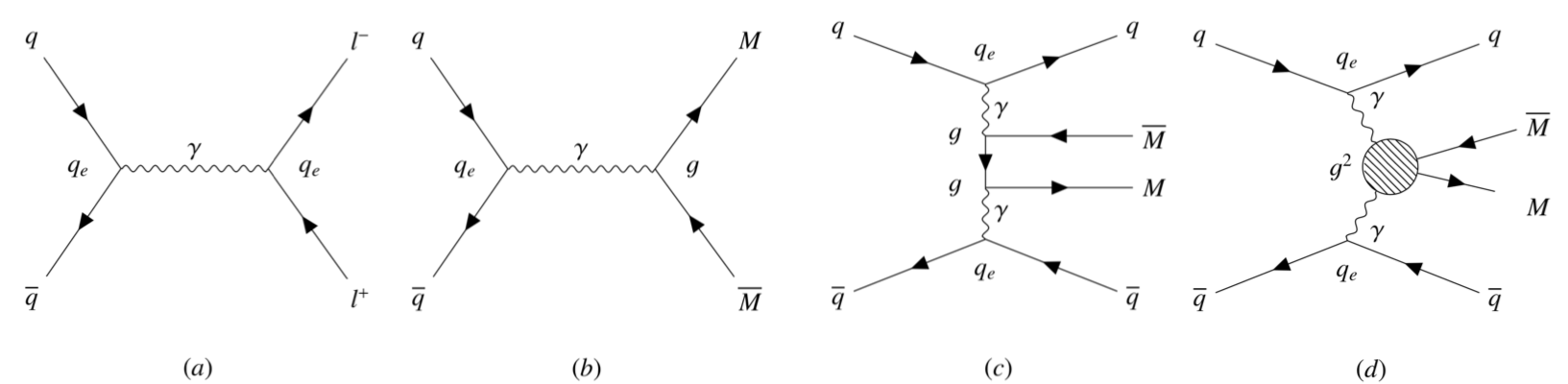}
\caption{Feynman-like tree-level graphs of (\textbf{a}): a Standard Model Drell--Yan (DY) process for lepton production from quark annihilation, with appropriate electric charges $q_{\rm e}$; (\textbf{b}) DY monopole--anti-monopole pair production from quark annihilation where $g$ is the monopole's magnetic charge; (\textbf{c}) monopole--anti-monopole pair production via photon-fusion (PF) (for monopole spins 0,$\frac{1}{2}$, and 1); (\textbf{d}) additional (contact) diagrams for monopole--anti-monopole pair production via PF (for monopole spins 0 and 1). The blob denotes the effective coupling. Wavy lines denote photons ($\gamma$), while continuous lines denote either fermions (quarks (q), antiquarks ($\overline {\rm q}$), and charged leptons ($l^{\pm})$) or monopole (anti-monopole) fields M ($\overline {\rm M}$) \cite{TikZ}.}
\label{fig:diagrams}
\end{figure}

Analytical predictions of kinematic distributions would serve as invaluable guides in monopole searches, such as those performed by the MoEDAL experiment at CERN, provided they fit within an acceptable field theory. But the non-perturbative nature of the coupling has also hindered a meaningful evaluation of scattering amplitudes in a quantum theory of monopoles. In the context of a dualised electromagnetic theory, with charge quantisation 
\begin{equation}\label{quatcond}
g q_e = \frac{1}{2}n(4\pi \epsilon_0 c)^{\xi}\hbar c, \;\; n\in Z
\end{equation}
the magnetic coupling $g$ is fixed as a large number, hence making the model non-perturbative. $c$ is the speed of light in vacuum, $\hbar$ is Planck's 
 constant, $\epsilon_0$ is the vacuum permittivity, $n$ is the linking number, and $\xi$ is 0 in CGS 
  Gaussian units and 1 in SI units. However, a perturbative description is recovered in the context of the low-energy effective field theory presented here and in the full publication \cite{fullreport}. As in all effective field theories (EFTs),  
   an effective coupling, in this case between the monopole and the photon, is proposed that adequately describes the physics in the low-energy limit. Motivated by arguments of classical scattering of monopoles off electrons~\cite{kMilton,scatt1,scatt3,scattlast}, it is proposed that the coupling is dependent on the Lorentz-invariant boost of the particle in the centre of mass frame, $\beta$. Clearly, monopole production described from this EFT is relevant only if these particles are produced at threshold where $\beta \ll 1$. In fact, for small enough $\beta$, this renders the effective coupling $g\beta$ perturbative. In this non-relativistic limit, a limit of relevance to the MoEDAL experiment at CERN \cite{MoEDAL} amongst others, the coupling becomes weak, and a perturbative theory is established. Hence, Feynman-like graphs as in Figure \ref{fig:diagrams} can be drawn within the context of this EFT only.

The total cross sections for monopole production by photon fusion were analytically derived by Kurochkin et al.~\cite{original1,original2}~for three different spin models, spins $0$, $\frac{1}{2}$, and $1$, using the dualised vertex amplitudes for scalar quantum electrodynamics, (SQED), 
 and the $e^-e^-\gamma$ and $W^+W^-\gamma$ Standard Model vertices. But these amplitudes are very specific to the Standard Model Lagrangian and are not transferable to a general theory of monopoles. Specifically, the spin $\frac{1}{2}$ particle is defined in a minimally coupling theory, mirroring the behaviour of the electron, with a magnetic moment generated through spin interactions (gyromagnetic ratio $g_{e}=1$), while the monopole with spin $1$ is assumed to behave as the $W$ boson, which gains a magnetic moment naturally through interactions generated in an electroweak theory with spontaneous symmetry breaking, and hence adopting a gyromagnetic ratio $g_{W}=2$ \cite{original1,original2}.   These assumptions cannot be made, and this calls for a more careful treatment in monopole model building.

In the context of a perturbative coupling, this work introduces a model-independent way of treating the magnetic moment, which influences amplitudes though a variable parameter $\kappa$. Using this construction, the kinematic distributions for monopole production by Drell--Yan (DY) and photon fusion (PF) processes, with diagrams drawn in Figure \ref{fig:diagrams}, are calculated. These are reduced to \cite{original1,original2} when the spin $\frac{1}{2}$ fermionic monopole takes $\kappa=0$ and the spin 1 monopole takes $\kappa=1$, mirroring the electron and $W$ boson, respectively. 

In Section~\ref{sec-1}, each spin model is treated analytically. A comparison of the three spin models is given in Section~\ref{sec-2} along with an assessment of their detectability at current colliders. Finally, the conclusion appears in Section~\ref{sec-3}.

\section{Analytical Calculations for Monopole Production Processes}
\label{sec-1}
In  dualised theory, the magnetic coupling $g$ is inversely proportional to the electric coupling $q_e$, as required by the quantisation condition (\ref{quatcond}), so that any process dependent on $g$ is non-perturbative. But the computation of scattering amplitudes involving monopoles requires the kinematics be confined to the perturbative regime of the theory. If the coupling is $\beta$-dependant, the theory can be treated perturbatively for monopoles produced at threshold. The coupling, and hence the magnetic structure constant $\alpha_g$, are defined as

\begin{align}\label{coupling}
	g(\beta)=g\beta^{\xi}, \;\;\;\;\; \alpha_g(g) = \frac{g^2(\beta)}{4\pi}, \;\;\;\;\; 	\beta=\sqrt{1-\frac{4M^2}{s}}\;\;,
	\vspace*{-\baselineskip}
\end{align}
where $\xi=1$ in a $\beta$-dependent model, and $\xi=0$ in the equivalent (non-perturbative) $\beta$-independent model. $M$ is the monopole's mass, and $s$ is the centre of mass energy. ($s_{q\overline{q}}$ is the centre of mass energy of two colliding quarks in a DY process, and $s_{\gamma\gamma}$ is that of two fusing photons in PF.) 

Studies of the classical (tree-level) scattering of charged particles off magnetic monopoles as in~\cite{kMilton,scatt1,scatt3,scattlast} motivated this  $\beta$-dependence and is described more elaborately in \cite{fullreport}. As monopoles are expected to have TeV 
scale masses, this classical (low $\beta$) limit is precisely the range relevant in current and future collider experiments.

The choice of field theory is dictated by the spin of the monopole \cite{fullreport}. The spin $0$ monopole theory is represented by a dualised massive SQED, 
 the spin $\frac{1}{2}$ theory, by dualised massive QED, and the spin $1$ monopole theory by a dualised $U(1)$ gauged Proca theory. The latter two models are augmented by the presence of spinor and bosonic magnetic moment terms, respectively, which scale with the unknown dimensionless phenomenological parameters $\tilde{\kappa}$ and $\kappa$, respectively \cite{fullreport}. $\tilde{\kappa}=0$ and $\kappa=1$ represent the only renormalisable models at threshold and reproduce the Standard Model like couplings for the electron in the fermionic model and the $W$ boson in the bosonic monopole model. 
Each model describes the propagation and interactions of a monopole of mass $M$ in a $U(1)$ gauge-invariant theory. The Lagrangian in each case gains a kinetic term for the gauge field represented by the square of the field strength tenor $F_{\mu\nu}=\partial_{\mu}\mathcal{A}_{\nu}-\partial_{\nu}\mathcal{A}_{\mu}$, a mass term for the monopole field, and a kinetic term for the monopole field $\mathcal{D}_{\mu}\Phi$, which contains a coupling to the gauge field through the covariant derivative $\mathcal{D}_{\mu}=\partial_{\mu}-ig(\beta)\mathcal{A}_{\mu}$. Details on each Lagrangian and its content can be found in the full paper \cite{fullreport}.
Staying in the confines of the perturbative regime at low $\beta$, vertex amplitudes for the DY and PF are extracted, and kinematic variables are calculated analytically, along with their~distributions. 

\subsection{The Spin 0 Monopole}
\label{par-1}
The kinematics for a scalar monopole of mass $M$ is already a well-studied topic (\cite{MoEDAL}, for example). It is mentioned here for completeness. 
This model generates a three- and a four-point vertex with respective amplitudes 

\begin{align}
\notag	 V^3_{\mu}=-ig(\beta) (p_{1}+p_{2})_{\mu}\;\;\; \text{and}  \;\;\; V^4_{\mu\nu}=2ig^{2}(\beta)g_{\mu\nu},
\end{align}
where $g_{\mu\nu}$ is the Minkowski metric and $p_i$ are monopole momenta. The kinematic distributions for monopole pair production are derived analytically (see \cite{fullreport} for details) for PF and DY processes.
\vspace{6pt}
\paragraph*{\bf Pair Production by Photon Fusion\newline}
The kinematic distribution and total cross section for pair production by PF are derived from the matrix amplitude, which combines the t-channel, u-channel, and seagull graphs, depicted between their parent quarks in Figure \ref{fig:diagrams}c,d.

\begin{figure}[H]
\centering
\includegraphics[scale=0.45]{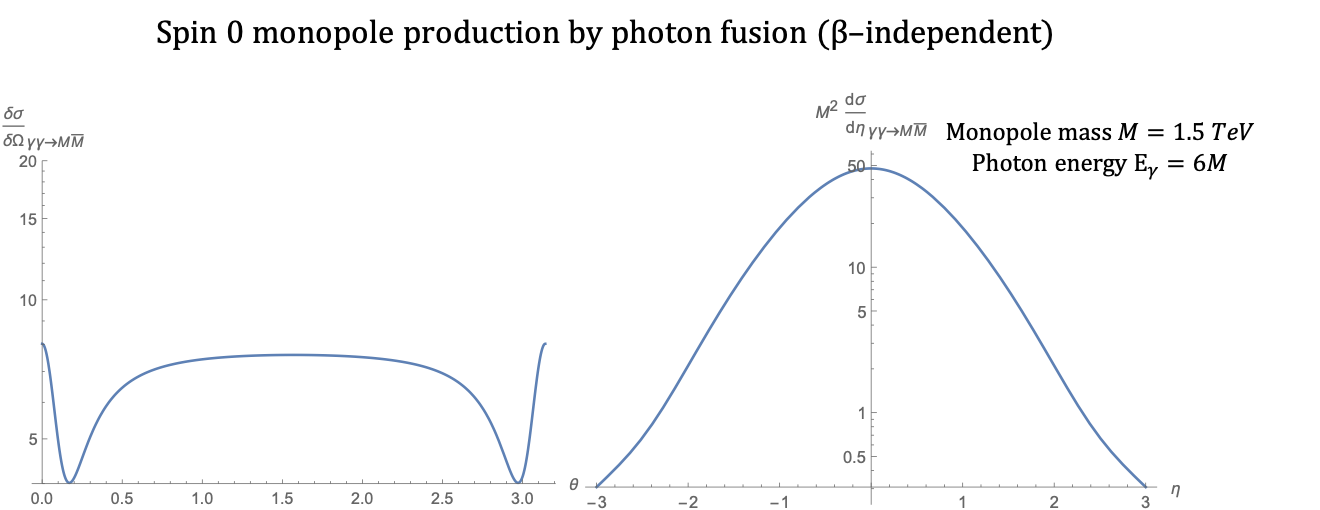}
\includegraphics[scale=0.4]{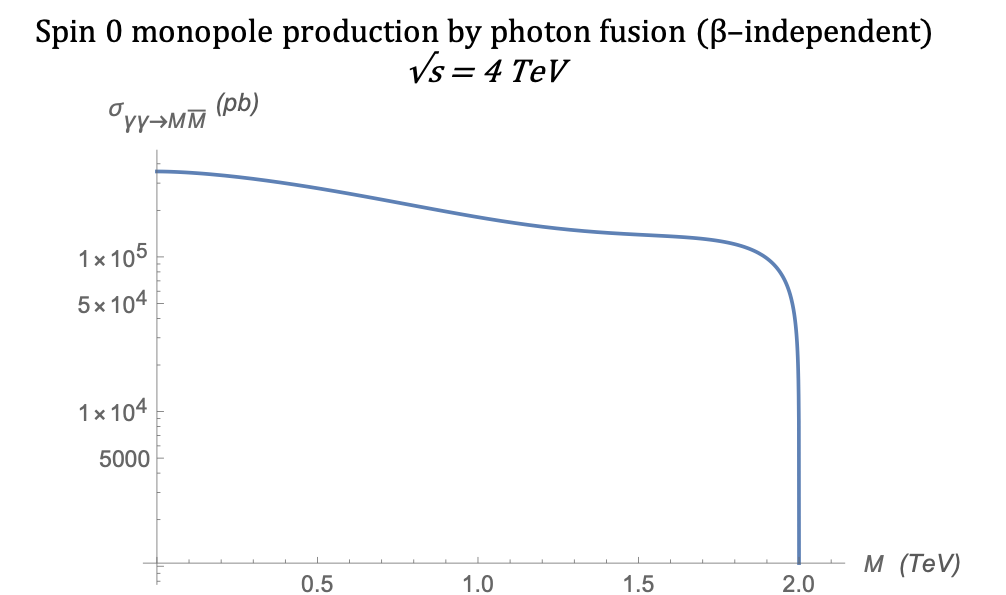}
\caption{Spin $0$ monopole production by PF: (\textbf{Left}) These plots show distributions for pair production in the centre of mass frame as functions of scattering angle $\theta$ and pseudo-rapidity $\eta$, which are focused in the central region. The monopoles have mass $M=1.5$ TeV and $\sqrt{\sgg}=2E_{\gamma}$, where $E_{\gamma}=6M$. (\textbf{Right}) The total cross section varies slowly with monopole mass $M$ at $\sqrt{\sgg}=4$ TeV until it drops off sharply in the kinematically forbidden region $M>\sqrt{\sgg}/2$.}\label{PFdsigmaSpin0}
\end{figure}

\begingroup\makeatletter\def\f@size{8}\check@mathfonts
\def\maketag@@@#1{\hbox{\m@th\fontsize{10}{10}\selectfont \normalfont#1}}

\begin{align}
	& \frac{d\sigma_{\gamma\gamma\rightarrow M\overline{M}}^{S=0}}{d\Omega} = \frac{\alpha_{g}^{2}(\beta)\beta}{2\sgg} \left\{1+\left[1-\left(\frac{2(1-\beta^{2})}{(1-\beta^{2}\cos^{2}\theta)}\right)\right]^{2}\right\},\;\;\;\; \sigma^{S=0}_{\gamma\gamma\rightarrow \overline{M}M} =\frac{4\pi\alpha_{g}^{2}(\beta)\beta}{\sgg}\left[2-\beta^{2}-\frac{1}{2\beta}(1-\beta^{4})\ln{\left(\frac{1+\beta}{1-\beta}\right)}\right],\label{dsigmadomegazeropf}
\end{align}
\endgroup
where $\beta= \Big(1-\frac{4M^2}{\sgg}\Big)^{\frac{1}{2}}$. The production is manifestly central. The integrated cross section agrees with~\cite{original1,original2} and is displayed graphically in Figure \ref{PFdsigmaSpin0}, as is the differential form in (\ref{dsigmadomegazeropf}) for a monopole with mass $M=1.5$ TeV at $\sqrt{\sgg}=2E_{\gamma}$, where $E_{\gamma}=6M$. The total cross section on the right of Figure  \ref{PFdsigmaSpin0} disappears in the kinematically forbidden region $M>\sqrt{\sgg}/2$, and the production is non-divergent.

\vspace{6pt}
\paragraph*{\bf Pair Production by Drell--Yan\newline}
The kinematic distributions for monopole production by Drell--Yan as drawn in Figure \ref{fig:diagrams}b are calculated assuming all quarks are massless.

\begin{figure}[H]
\centering
\includegraphics[scale=0.43]{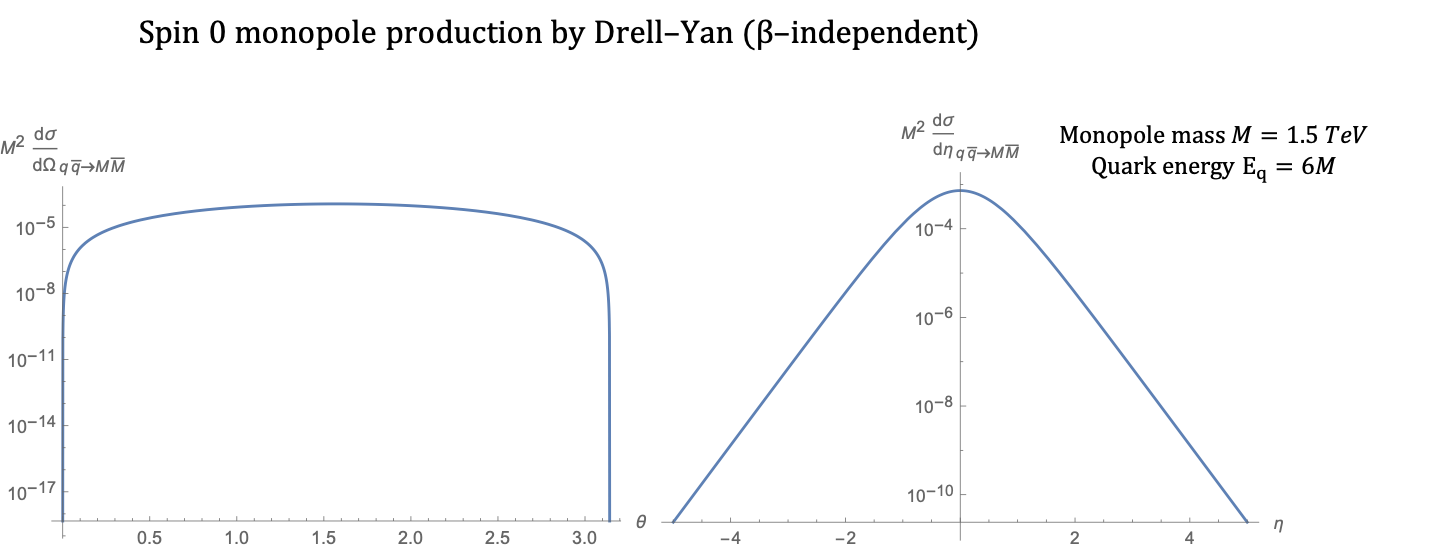}
\includegraphics[scale=0.39]{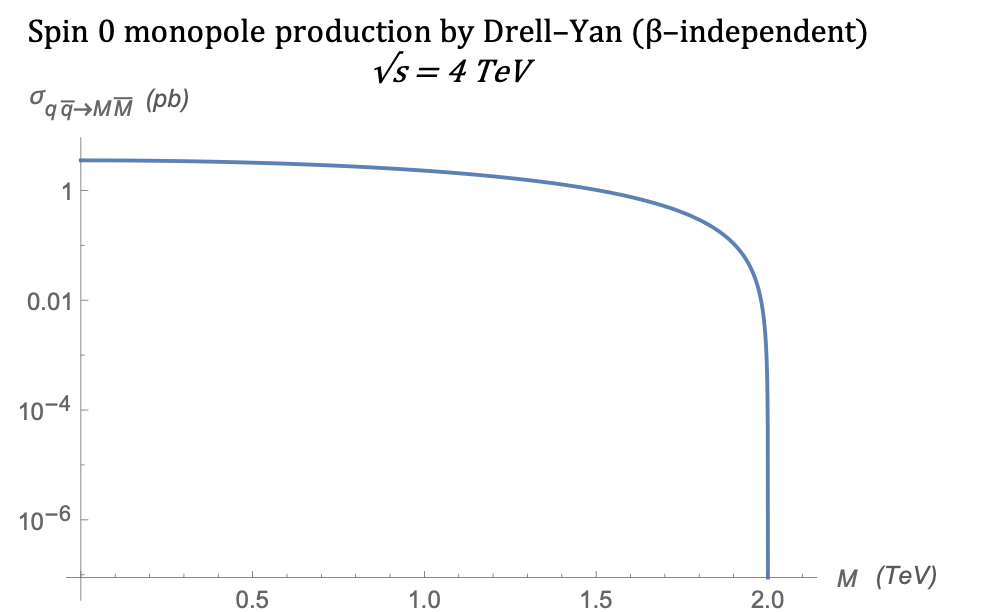}
\caption{Spin $0$ monopole production by Drell--Yan (DY): (\textbf{Left}) The figure shows that the production from massless quarks, with $M=1.5$ TeV and $\sqrt{\sqq}=2E_{q}$ for $E_{q}=6M$, is predominantly concentrated in the central region. (\textbf{Right}) The total cross section for pair production in dualised SQED 
	 is finite, as shown for $\sqrt{\sqq}=4$ TeV, in the same way as PF 
	 production was.}\label{DYdsigmadetaSpin0}
\end{figure}

\begingroup\makeatletter\def\f@size{9}\check@mathfonts
\def\maketag@@@#1{\hbox{\m@th\fontsize{10}{10}\selectfont \normalfont#1}}
 \begin{align}\label{xsecdeff}
	\frac{d\sigma_{q\overline{q}\rightarrow M\overline{M}}^{S=0}}{d\Omega} = \frac{5\alpha_{g}(\beta)\alpha_{e}}{72 \, \sqq}\beta^3(1-\cos^2(\theta)) \;\;\;\;\;\;\;\;\; \text{and} \;\;\;\;\;\;\;\;\;
 \sigma_{q\overline{q}\rightarrow M\overline{M}}^{S=0} = \frac{5\pi \alpha_{g}(\beta)\, \alpha_{e}}{27 \, \sqq}\, \beta^3,
\end{align}
\endgroup
where $\beta= \Big(1-\frac{4M^2}{\sqq}\Big)^{\frac{1}{2}}$.
This last expression, as in all subsequent DY cases, is valid in experiments with particle--anti-particle bunch crossings as in the case of the Tevatron and is doubled when using a symmetric beam experiment such as the Large Hadron Collider (LHC). Equations (\ref{xsecdeff}) are drawn in Figure \ref{DYdsigmadetaSpin0} for a monopole with mass $M=1.5$ TeV at $\sqrt{\sgg}=2E_{\gamma}$, where $E_{\gamma}=6M$. The production is even more central that in the PF
 case, and the total cross section is non-divergent.

\subsection{The Spin $\frac{1}{2}$ Monopole}
\label{par-2}
\begin{figure}[H]
\centering
\includegraphics[scale=0.43]{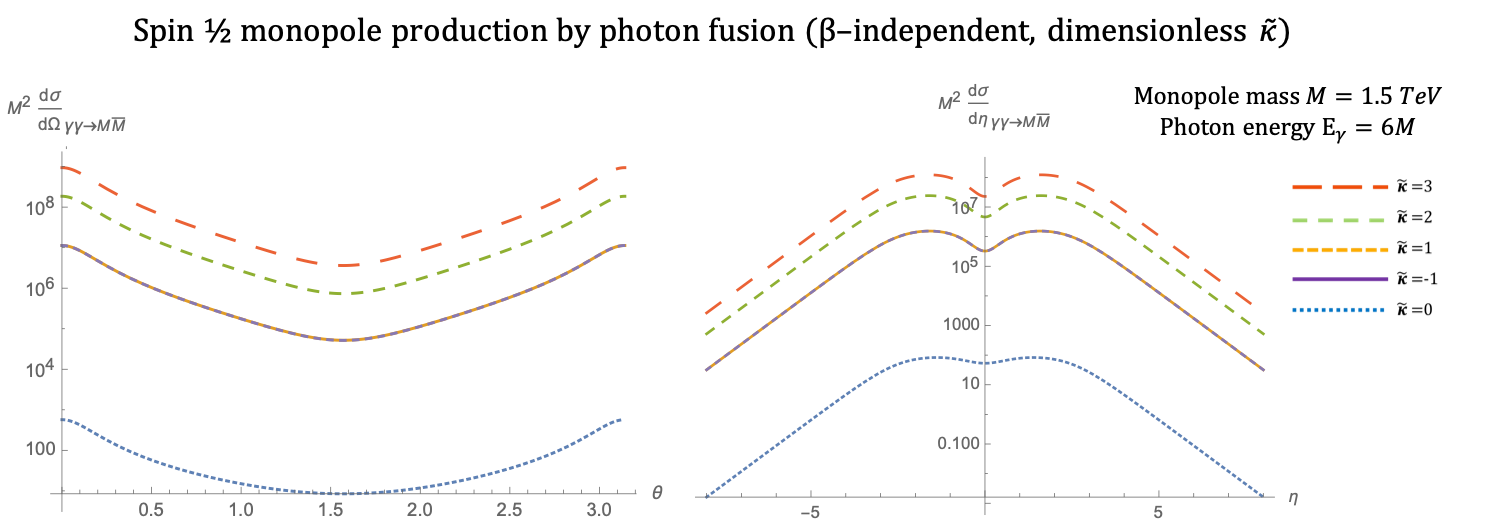}
\includegraphics[scale=0.38]{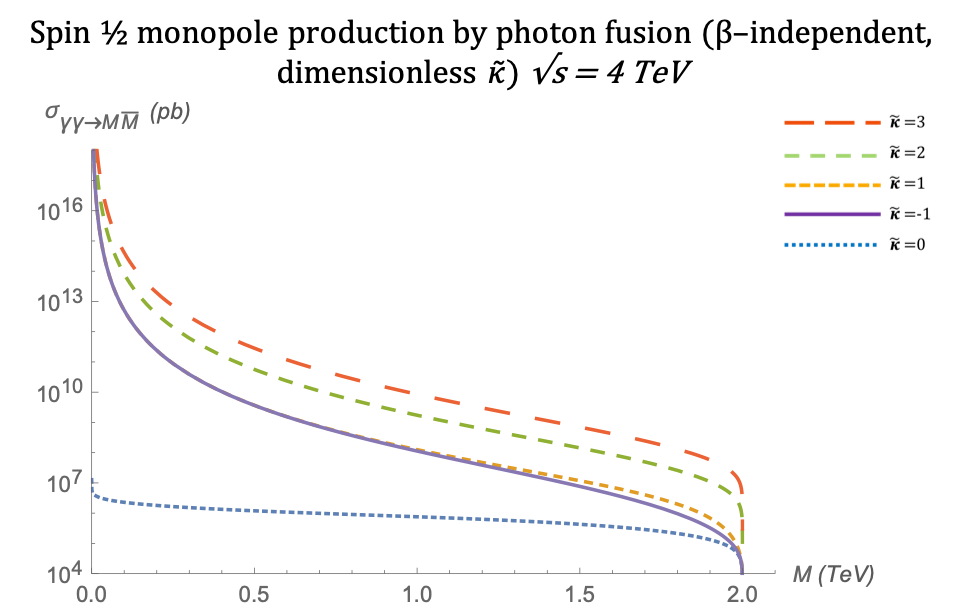}
\caption{Spin $\frac{1}{2}$ monopole production by PF: (Left) For $M=1.5$ TeV and $E_q=6M$, as $\tilde{\kappa}$ changes, the distributions change only by a scaling factor, and production is concentrated away from the central axis. (This contrast with the s=$0$ case is expected.) The $\tilde{\kappa}=0$, representing dualised QED, is unique as the only renormalisable, unitary case. (Right) For various values of $\tilde{\kappa}\neq0$, the total cross section at $\sqrt{\sqq}=4$ TeV diverges as $M\rightarrow0$, where the monopole  becomes non-relativistic.}\label{LeeYangDsigDstuff12label}
\end{figure}

As stated in Section 2, the Lagrangian for the spinor monopole includes a moment term that scales with $\tilde{\kappa}$ \cite{fullreport}. This parameter can be constrained through measurements of the magnetic moment of the monopole, which now has a gyromagnetic ratio of $g_R=2(1+2\tilde{\kappa})$, but also through the only vertex amplitude coupling the photon to monopoles.

\begin{align} \label{vertexSpinor}
	V^3_{\mu}=-ig(\beta) \left(\gamma_{\mu}+\frac{1}{2M}\tilde{\kappa} k^{\sigma}[\gamma_{\sigma},\gamma_{\mu}]\right),
\end{align}
where $k^{\sigma}$ is the photon momentum and $g_{\mu\nu}$ is the Minkowski metric. Notice that the second term shows explicitly that this effective field theory is non-renormalisable at scales $k^2>M^2$. The mass-dependance in the vertex amplitude is required on dimensional grounds. 
 \vspace{6pt}
\paragraph*{\bf Pair Production by Photon Fusion\newline}
Having vertex (\ref{vertexSpinor}) only, spinor monopole pair production only has t- and u-channel contributions, as depicted emanating from quark lines in Figure \ref{fig:diagrams}c. The $\tilde{\kappa}$-dependent differential cross section is 
\begin{align}\label{dxsec12}
\begin{split}
 \frac{d\sigma^{S=\frac{1}{2}}_{\gamma\gamma\rightarrow M\overline{M}}}{d\Omega} &=\frac{  \alpha_g^2(\beta) \beta}{4 \sgg (\beta^2 \cos ^2(\theta )-1)^2}(-\beta^6 \kappa ^4 \sgg^2 \cos ^6(\theta )-2 \beta^4 (\kappa ^4 \sgg^2+4)\\&+\beta^2 (48 \kappa  \sqrt{\sgg-\beta^2 \sgg}+2 \kappa ^4 \sgg^2+32 \kappa ^2 \sgg+8)-\beta^4 \cos ^4(\theta ) ((2 \beta^2+3) \kappa ^4 \sgg^2\\&+8 \kappa ^2 \sgg+4)+\beta^2 \cos ^2(\theta ) (2 \beta^4 \kappa ^4 \sgg^2+8 \beta^2 (5 \kappa ^2 \sgg+1)-48 \kappa  \sqrt{\sgg-\beta^2 \sgg}\\&+3 \kappa ^4 \sgg^2-60 \kappa ^2 \sgg-8)+(\kappa ^2 \sgg-2)^2),
\end{split}
\end{align}
where $\beta=\sqrt{1-\frac{4M^2}{\sgg}}$. For $\kappa=0$, Standard Model dual QED is recovered and the renormalisability (finite cross section in the $\sgg\rightarrow \infty$ limit) is restored. Figure \ref{LeeYangDsigDstuff12label} shows a scaling of distributions with $\tilde{\kappa}$ and a degeneracy between positive and negative $\tilde{\kappa}$.
The total cross section is 
\begingroup\makeatletter\def\f@size{9}\check@mathfonts
\def\maketag@@@#1{\hbox{\m@th\fontsize{10}{10}\selectfont \normalfont#1}}
\begin{align}
\begin{split}
	 \sigma^{S=\frac{1}{2}}_{\gamma\gamma\rightarrow M\overline{M}}&=\frac{\pi  \alpha_g^2(\beta)}{3\sgg} \Bigg(3 \beta^4 \kappa ^4 \sgg^2 \ln\Big(\frac{1-\beta}{1+\beta}\Big)+6 \beta^4 \ln \Big(\frac{1-\beta}{1+\beta}\Big)-7 \beta^3 \kappa ^4 \sgg^2+12 \beta^3-6 \beta^2 \kappa ^4 \sgg^2 \ln\Big (\frac{1-\beta}{1+\beta}\Big)\\&+6 \beta^2 \kappa ^2 \sgg \ln \Big(\frac{1-\beta}{1+\beta}\Big)-72 \beta \kappa  \sqrt{-(\beta^2-1) \sgg}-36 \beta^2 \kappa  \sqrt{-(\beta^2-1) \sgg} \ln\Big(\frac{1-\beta}{1+\beta}\Big)\\&-36 \kappa  \sqrt{-(\beta^2-1) \sgg} \ln \Big(\frac{1-\beta}{1+\beta}\Big)-15 \beta \kappa ^4 \sgg^2-9 \kappa ^4 \sgg^2 \ln\Big(\frac{1-\beta}{1+\beta}\Big)\\&-132 \beta \kappa ^2 \sgg-60 \kappa ^2 \sgg \ln \Big(\frac{1-\beta}{1+\beta}\Big)-24 \beta-18 \ln \Big(\frac{1-\beta}{1+\beta}\Big)\Bigg).\label{totxsec12}
\end{split}
\end{align}
\endgroup
Setting $\kappa=0$, expression (\ref{totxsec12}) reduces to that given in the literature (\cite{original1,original2} for example). As seen in Figure~\ref{LeeYangDsigDstuff12label}, the $\tilde{\kappa}=0$ case remains the only unitary option in the $\sgg\rightarrow\infty$ limit. Equation (\ref{totxsec12}) also diverges as $M\rightarrow0$ (relativistic monopole) outside the perturbative regime for $\tilde{\kappa}\neq0$.

\begin{figure}[H]
\centering
\includegraphics[scale=0.4]{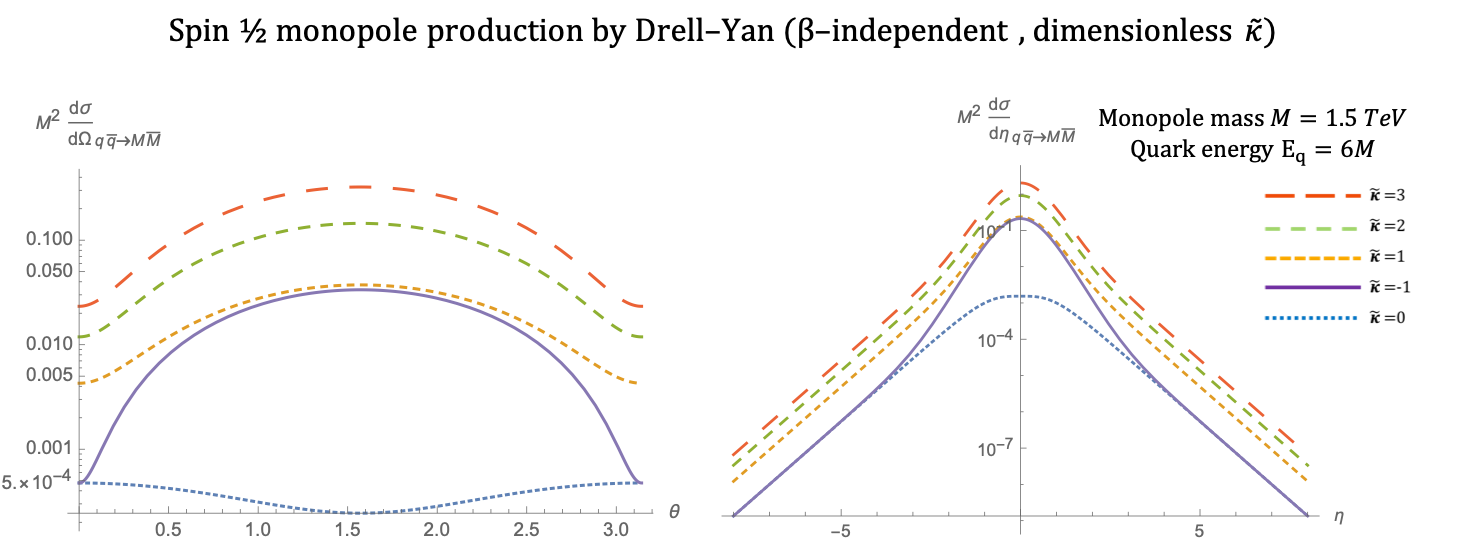}
\includegraphics[scale=0.38]{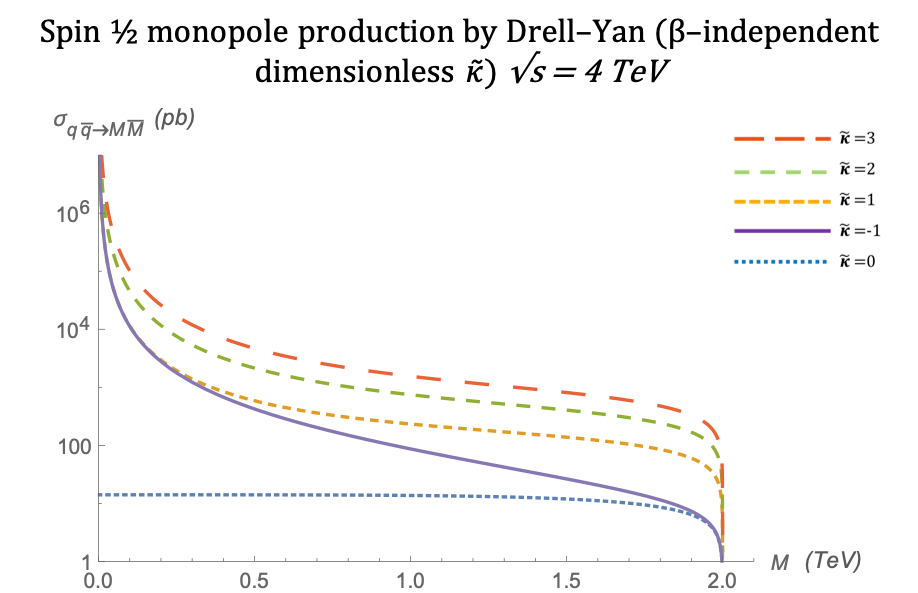}
\caption{Spin $\frac{1}{2}$ monopole production by DY: (Left) The angular and rapidity distributions for various values of the parameter $\tilde{\kappa}$ demonstrate rather more contrasting behaviours between the $\tilde{\kappa}$ cases, unlike the PF distributions, and also show a much more central production. Here, the monopole mass is $M=1.5$ TeV, and the quark energy is $E_q=6M$. (Right) For various values of $\tilde{\kappa}\neq0$, the total cross section at $\sqrt{\sqq}=4$ TeV diverges as $M\rightarrow0$, where the monopole enters a non-relativistic regime.}\label{LeeYangDsigDstuff12labeldy}
\end{figure}

\vspace{6pt}
\paragraph*{\bf Pair Production by Drell--Yan\newline}
The differential cross section distribution for fermionic monopole production by DY is represented by an s-channel graph of the type shown in Figure \ref{fig:diagrams}b, where $\beta=\sqrt{1-\frac{4M^2}{\sqq}}$. Analytically,

\begingroup\makeatletter\def\f@size{9.5}\check@mathfonts
\def\maketag@@@#1{\hbox{\m@th\fontsize{10}{10}\selectfont \normalfont#1}}

\begin{align}\label{Xsecdeffspin12}
	\frac{d\sigma^{S=\frac{1}{2}}_{q\overline{q}\rightarrow M\overline{M}}}{d\Omega} &=\frac{5 \alpha_e \alpha_g(\beta) }{36 \sqq}\Big(\beta^3 (\cos ^2(\theta )-\kappa ^2 \sqq \cos ^2(\theta )-\kappa ^2 \sqq-1)+\beta(4 \kappa  \sqrt{\sqq-\beta^2 \sqq}+2 \kappa^2 \sqq+2)\Big)
\end{align}
\endgroup
in the massless quark limit, and total cross section is 
\begin{equation}\label{thisone}
\sigma^{S=\frac{1}{2}}_{q\overline{q}\rightarrow M\overline{M}}=\frac{10 \pi  \beta \alpha _e \alpha _g(\beta) }{27 \,\sqq}\left(3-\beta^2-(2 \beta ^2-3) \kappa ^2 \sqq+6 \kappa  \sqrt{\sqq-\beta ^2 \sqq}\right).
\end{equation}
Equation (\ref{Xsecdeffspin12}) happens to have a unitary behaviour, converging as $\sqq\rightarrow\infty$ for all $\tilde{\kappa}$ drawn on the left in Figure \ref{LeeYangDsigDstuff12labeldy}. The production diverges as $M\rightarrow0$ outside the perturbative regime, however, as shown for $\tilde{\kappa}\neq0$ by Equation (\ref{thisone}), drawn on the right in Figure \ref{LeeYangDsigDstuff12labeldy}.

\subsection{The Spin 1 Monopole}
\label{par-3}
Finally, the Lagrangian for the spin-$1$ monopole of mass $M$ in a dualised gauge theory draws from the Lee--Yan Lagrangian \cite{LeeYang} and is further extended to include the magnetic moment term proportional to a dimensionless $\kappa$ which, could be constrained by  magnetic moment measurements as it contributes to the gyromagnetic ratio $g_R=1+\kappa$. The three- and four-point vertex amplitudes are also $\kappa$-dependent

\begin{align}\label{spin1feyngraphs}
\begin{split}
	V^3_{\mu\nu\rho}=& -ig(\beta)\left(-g^{\nu\mu}(-\kappa p_2+\kappa p_1+p_1)^{\rho}-g^{\mu\rho}(p_2+\kappa p_2- \kappa p_1)^{\nu}+g^{\rho\nu}(p_1+p_2)^{\mu}\right)\;,\\
	V^4_{\mu\nu\sigma\rho}=& -2ig(\beta)^{2}(g^{\mu\nu}g^{\sigma\rho})+ig(\beta)^{2}(g^{\mu\sigma}g^{\nu\rho}+g^{\mu\rho}g^{\nu\sigma}) \;,
\end{split}
\end{align}
where $g_{\mu\nu}$ is the Minkowski metric and $p_i$ are monopole momenta.
\vspace{6pt}
\paragraph*{\bf Pair Production by Photon Fusion\newline}
The amplitude for spin $1$ monopole pair production by PF carries contributions from a t-channel, a u-channel, and a seagull graph, shown between parent quark lines in Figure \ref{fig:diagrams}c,d. The differential and total cross section distributions are calculated analytically for these amplitudes and are given by

\begin{figure}[H]
\centering
\includegraphics[scale=0.42]{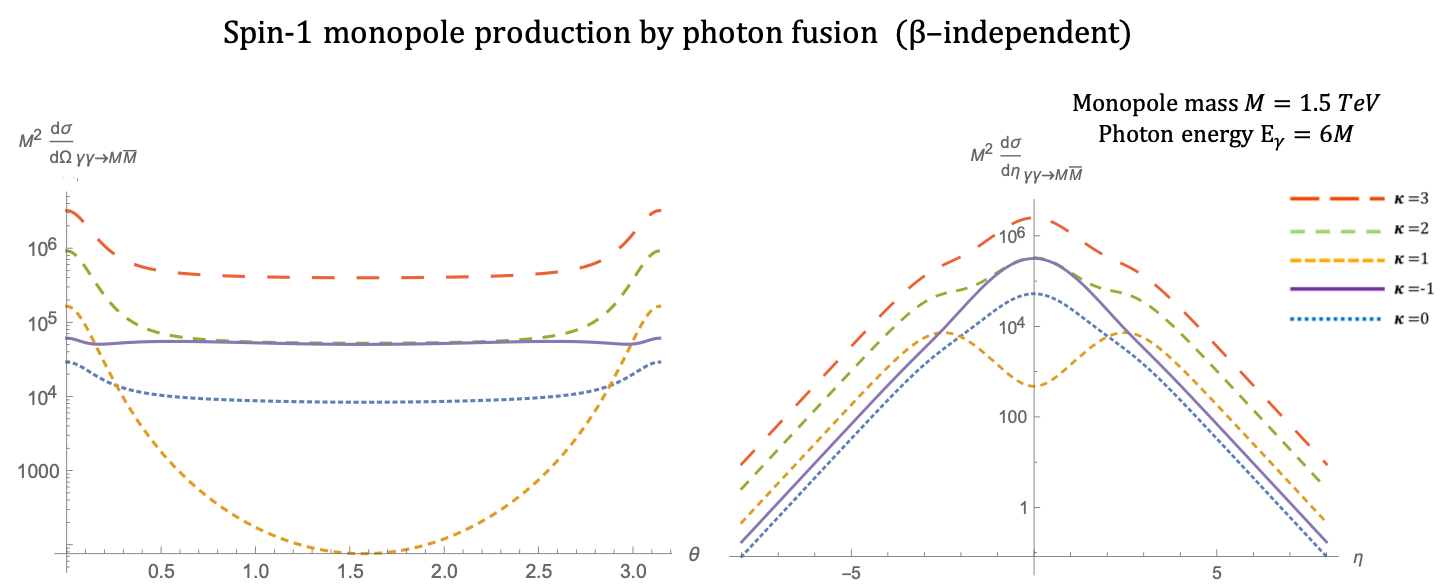}
\includegraphics[scale=0.33]{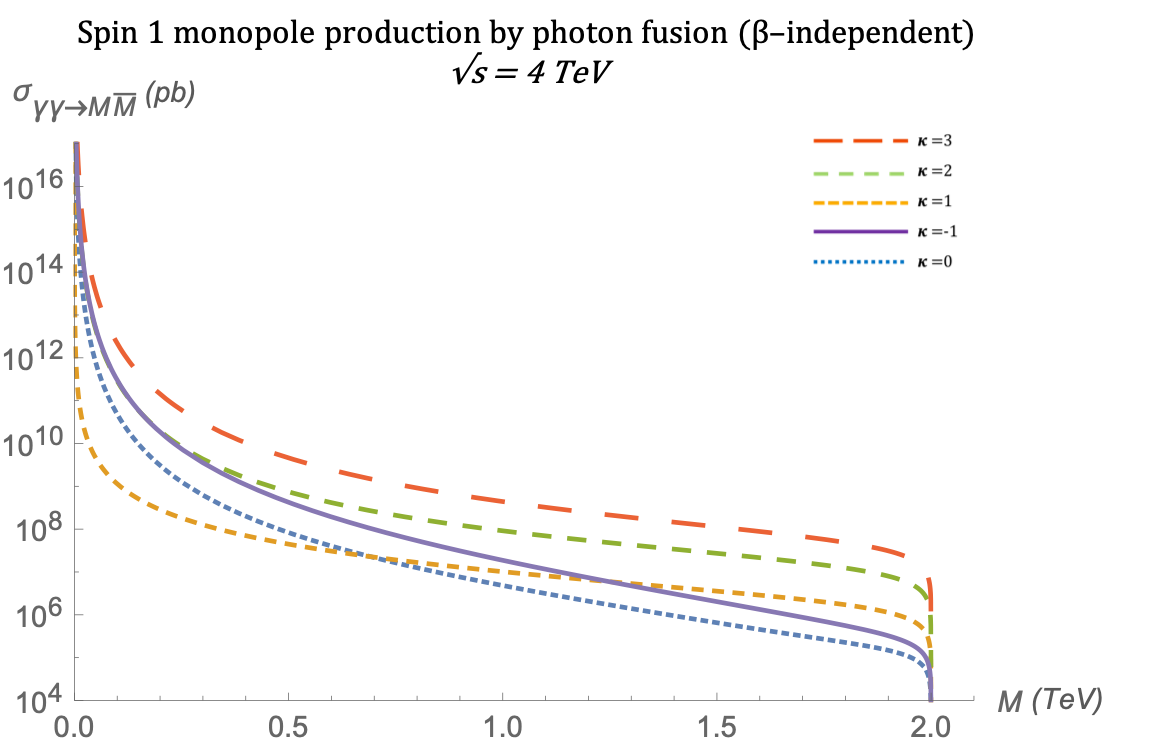}
\caption{Spin $1$ monopole production by PF: (\textbf{Left}) For different values of $\kappa$, at $M=1.5$ TeV, $\sqrt{\sgg}=2E_{\gamma}$, and  $E_{\gamma}=6M$, the $\kappa=1$ distributions are uniquely unitary, showing a depression of the cross section in the central region. $\theta$ is the scattering angle, and $\eta$ is pseudo-rapidity. (\textbf{Right}) The cross section for all $\kappa$, at $\sqrt{s_{\gamma\gamma}}=4$ TeV, diverges as $M\rightarrow0$, where the monopole naturally becomes non-relativistic.}\label{LeeYangDsigDomega1pflabel}
\end{figure}

\begingroup\makeatletter\def\f@size{8}\check@mathfonts
\def\maketag@@@#1{\hbox{\m@th\fontsize{10}{10}\selectfont \normalfont#1}}

\begin{align}
\begin{split}\label{dxsec1k}
\frac{d\sigma^{S=1}_{\gamma\gamma\rightarrow M\overline{M}}}{d\Omega}=&\frac{\alpha_g^2(\beta) \beta}{16 \left(\beta^2-1\right)^2 \sgg \left(\beta^2 \cos ^2(\theta )-1\right)^2} \Bigg(48 \beta^8+\beta^6 (\kappa-1)^4 \cos ^6(\theta )\\&-144 \beta^6+2 \beta^4 \left(3 \kappa^4+28 \kappa^3+42 \kappa^2-4 \kappa+79\right)-2 \beta^2 \left(11 \kappa^4+60 \kappa^3+58 \kappa^2+12 \kappa+35\right)\\&+\beta^4 \left(24 \beta^4+2 \beta^2 \left(\kappa^4+12 \kappa^3-10\kappa^2-20 \kappa-7\right)+9 \kappa^4-36 \kappa^3+22 \kappa^2+28 \kappa+1\right) \cos ^4(\theta )\\&-\beta^2 \bigg(48 \beta^6+2 \beta^4 \left(\kappa^4+4 \kappa^3-34 \kappa^2-28 \kappa-55\right)-4 \beta^2 \left(3 \kappa^4-42 \kappa^2-8 \kappa-29\right)\\&+35 \kappa^4-44 \kappa^3-78 \kappa^2-12 \kappa-29\bigg) \cos ^2(\theta )+29 \kappa^4+44 \kappa^3+46 \kappa^2+12 \kappa+21\Bigg)  \; ,
\end{split}\\
\begin{split}\label{totxsec1k}
\sigma_{\gamma\gamma\rightarrow M\overline{M}}^{S=1}&=\frac{\pi  \alpha^2_g(\beta)}{12 (\beta^2-1)^2 \sgg} \Bigg(-72 \beta^7+288 \beta^5-\beta^3 (-\kappa ^4+4 \kappa ^3+282 \kappa ^2+196 \kappa +263)\\&+6 (\beta^2-1) (6 \beta^6-6 \beta^4+\beta^2 (\kappa ^4+8 \kappa ^3+2 \kappa ^2-8 \kappa -9)-4 \kappa ^4-16 \kappa ^3+16 \kappa ^2+8 \kappa +2) \ln \left(\frac{1+\beta}{1-\beta}\right)\\& +3\beta (13 \kappa ^4-20 \kappa ^3+110 \kappa ^2+44 \kappa +29)\Bigg), 
\end{split}
\end{align}
\endgroup
respectfully, where $\beta=\sqrt{1-\frac{4M^2}{\sgg}}$, and are shown graphically in Figure \ref{LeeYangDsigDomega1pflabel}. While the differential forms diverge for $\kappa\neq1$ as $\sgg\rightarrow\infty$, the total cross section has a power-law divergence as $M\rightarrow0$, where the perturbative argument is lost.

\vspace{6pt}
\paragraph*{\bf Pair Production by Drell--Yan\newline}
Last but not least, the kinematic distributions for monopole pair production by the s-channel interaction, as in Figure \ref{fig:diagrams}b, are drawn using the analytical expressions in the massless quark limit

\begin{figure}[H]
\centering
\includegraphics[scale=0.4]{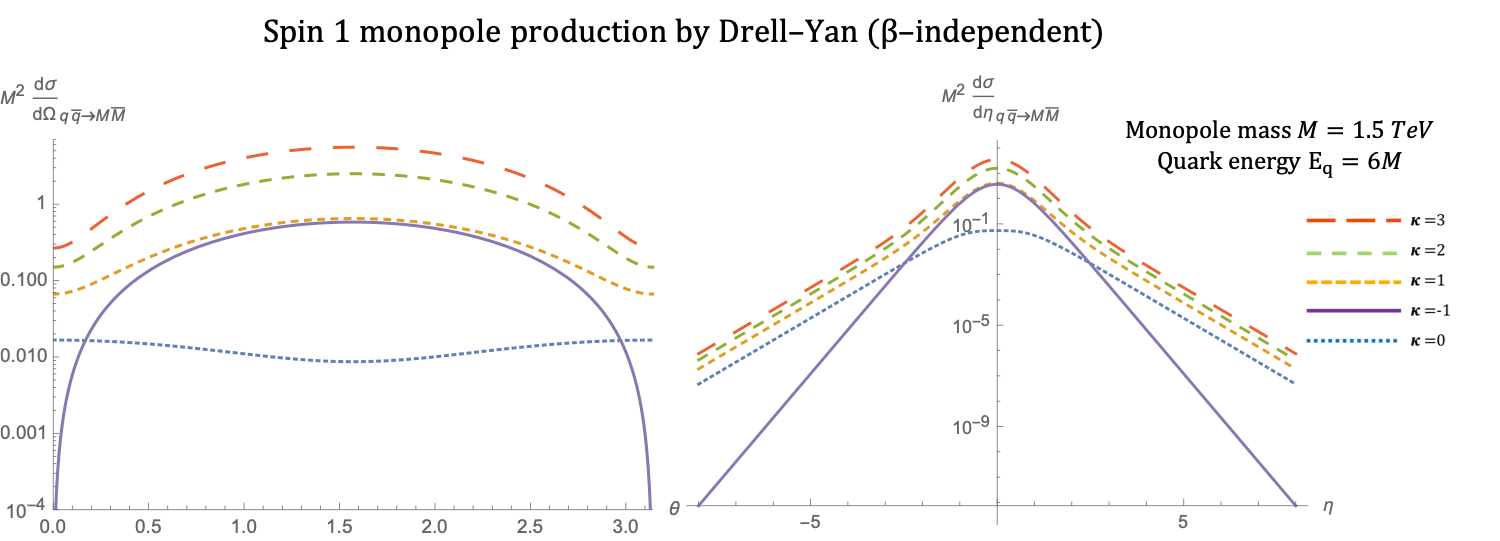}
\includegraphics[scale=0.38]{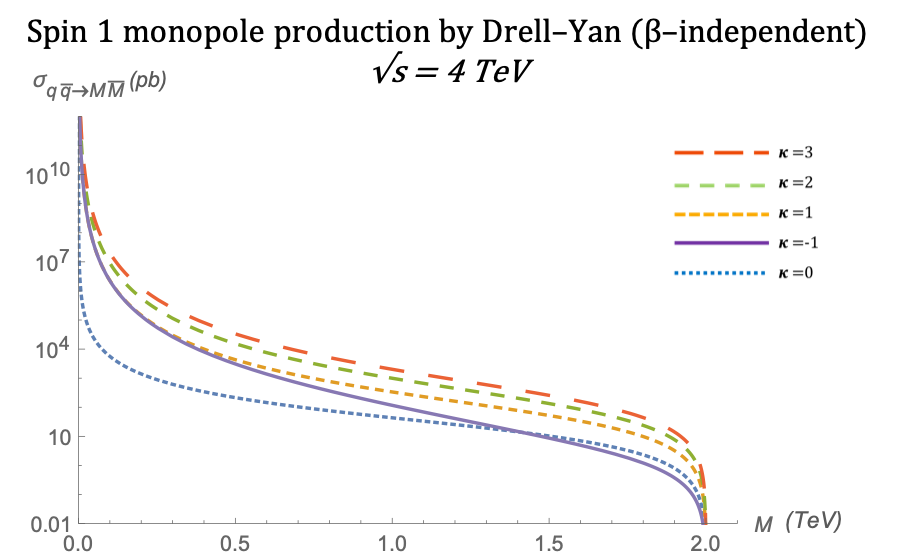}
\caption{Spin 1 monopole production by DY: (\textbf{Left}) In the massless quarks limit, with $M=1.5$ TeV and $\sqrt{\sqq}=2E_q$ at $E_q=6M$, the value of $\kappa$ influences the behaviour of the distribution but does not on its own reflect the unitarity of the model in the $\kappa=1$ case. (\textbf{Right}) The cross section for all $\kappa$, at $\sqrt{\sqq}=4$~TeV, diverges as $M\rightarrow0$, where the monopole naturally becomes non-relativistic.}\label{LeeYangDsigDomega1label}
\end{figure}

\begingroup\makeatletter\def\f@size{8}\check@mathfonts
\def\maketag@@@#1{\hbox{\m@th\fontsize{10}{10}\selectfont \normalfont#1}}

\begin{align}\label{Xsecdeffspin1}
	\frac{d\sigma^{S=1}_{q\overline{q}\rightarrow M\overline{M}}}{d\Omega}&=\frac{5 \beta ^3 \alpha _e \alpha _g(\beta)}{288 \left(\beta ^2-1\right) M^2}\Bigg(3\beta^4\big(\cos^2 \theta-1\big)+\beta^2\big(2\kappa^2(\cos^2 \theta+1)+8\kappa-4\cos^2 \theta+8\big) \nonumber \\
				& +2\kappa^2(\cos^2 \theta-3)-8\kappa+\cos^2 \theta-5\Bigg)\;,
\end{align}
\begin{equation}\label{totxsecdyk}
 \sigma^{S=1}_{q\overline{q}\rightarrow M\overline{M}}=\frac{5 \pi \sqq \alpha _e \alpha _g(\beta)}{432 M^4}\Big(1-\frac{4M^2}{\sqq} \Big)^{\frac{3}{2}}\left( 8\kappa^2-(4\kappa^2+12\kappa+10)  \Big(1-\frac{4M^2}{\sqq} \Big)+12\kappa+3\Big(1-\frac{4M^2}{\sqq} \Big)^2+7 \right)\;,
\end{equation}
\endgroup
where $\beta=\sqrt{1-\frac{4M^2}{\sqq}}$. They are plotted in Figure \ref{LeeYangDsigDomega1label}. As an isolated process, the cross section distribution converges as $\sqq\rightarrow\infty$ only for $\kappa=0$. The unitarity of the model for $\kappa=1$ is expected for the SM 
W boson once all contributing s-channel electroweak processes are included, where the W and Z bosons contribute as intermediate virtual states. This is, of course, not guaranteed for the monopole, which does not couple to the W and Z bosons. Hence, the non-unitary behaviour is not surprising. The total cross section also has a power-law divergence as $M\rightarrow0$, as perturbation theory breaks down.

\section{A Comparison of the Total Cross Sections and Small Coupling Limits}
\label{sec-2}

\begin{figure}[H]
\centering
\includegraphics[scale=0.35]{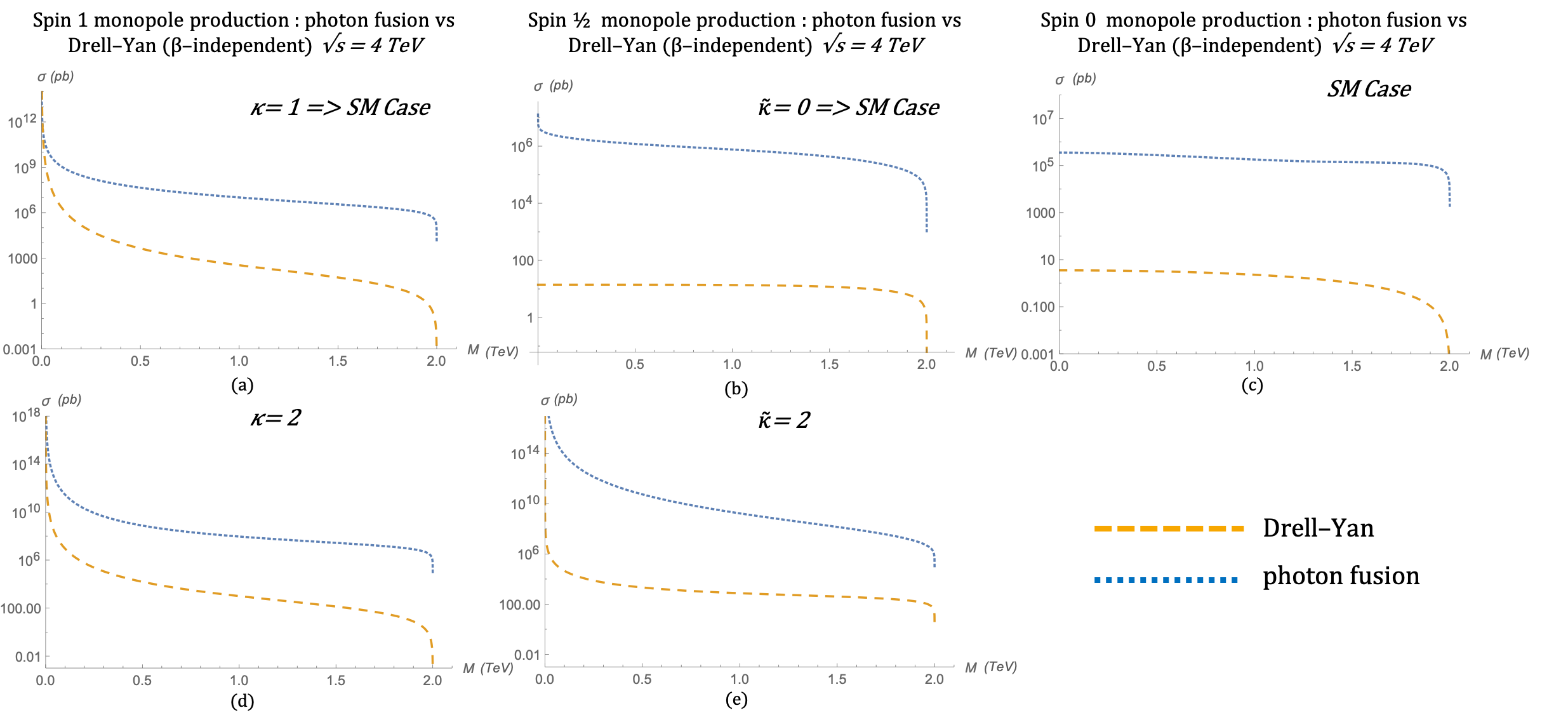}\centering
\caption{The production cross sections for PF decidedly dwarf those for DY at $\sqrt{s_{qq/\gamma\gamma}}=4$ TeV. Shown here are (\textbf{a}) $s=1$ monopole production in the $\kappa=1$ (SM-like) case; (\textbf{b}) the $s=\frac{1}{2}$ case for $\tilde{\kappa}=0$; (\textbf{c}) the only  $s=0$ case, which has no magnetic moment; (\textbf{d}) the $s=1$  monopole cross section with $\kappa=2$; and (\textbf{e}) the $s=\frac{1}{2}$ monopole cross section with $\tilde{\kappa}=2$. }\label{PFvsDY}
\end{figure}

At interaction energies relevant to colliders, such as $\sqrt{s}=4$ TeV, PF dominates DY production by a long shot (cf. Figure \ref{PFvsDY}), independently of the value of $\kappa (\tilde{\kappa})$, as demonstrated in both the unitary (Figure~\ref{PFvsDY}a--c) and arbitrarily chosen non-unitary (Figure~\ref{PFvsDY}d,e) cases. 
As already discussed, the perturbative treatment is valid only for threshold pair production, $\beta<<1$. However, this limit has the additional setback that it renders the production cross sections for $s=1,\frac{1}{2}$ negligible for LHC-type experiments, for both DY and PF processes, in the well behaved cases, $\kappa(\tilde{\kappa})=1(0)$. This changes if the moment parameters are allowed to be very large, $\kappa(\tilde{\kappa})>>1$, even while the derivative magnetic couplings in (\ref{vertexSpinor}) and (\ref{spin1feyngraphs}) are forced to remain perturbative overall (see \cite{fullreport} for details). For a monopole momentum of order $M\beta$, this means

\begin{align}\label{limitcoupling}
g\kappa'\beta^2<1, \;\;\;\;\;\; \kappa'=\kappa(\tilde{\kappa}) \text{ for } s=1\left(1/2\right).
\end{align}
Then, after imposing a good infra-red behaviour as $\beta\rightarrow0$ and choosing a parameterisation that satisfies (\ref{limitcoupling}) trivially, $(\kappa'\beta g)^4\beta \stackrel{\stackrel{\beta\rightarrow0}{\kappa'\rightarrow\infty}}{=} |c_1|$ for some constant $c_1$, the cross sections for PF become finite non-negligible in the $\beta\rightarrow0$ limit, towering over the still-trivial DY in both non-zero spin models.
\begin{eqnarray}\label{totsec12lim}
\sigma^{S=\frac{1}{2}}_{\gamma\gamma\rightarrow M\overline{M}} \sim \frac{(\tilde \kappa \, g \, \beta)^4  \, \beta }{16\, \pi\, M^4} \, s \stackrel{\stackrel{{\beta \to 0}}{\tilde \kappa \to \infty}}{=} {\rm finite}, & \;\;\;\;\;\;\;\;\;\;
\sigma^{S=1}_{\gamma\gamma\rightarrow M\overline{M}}& \stackrel{\stackrel{\beta \to 0}{\kappa \to \infty}}{=}
\frac{29\, c_1}{64\, \pi \, s} = \; {\rm finite},
\nonumber \\
\sigma^{S=\frac{1}{2}}_{q\overline{q}\rightarrow M\overline{M}} \sim \frac{5 \, \alpha_e}{18\, M^2}  \,
(\tilde \kappa \, \beta \, g )^2 \, \beta  \, \stackrel{\stackrel{\beta \to 0}{\kappa \to \infty}}{\to} \, 0, & \;\;\;\;\;\;\;\;\;\;
\sigma^{S=1}_{q\overline{q}\rightarrow M\overline{M}} &\stackrel{\stackrel{\beta \to 0}{\kappa \to \infty}}{=}
\alpha_e \, \frac{10\, \sqrt{|c_1|}}{27\,s} \, \beta ^{\frac{5}{2}} \quad \stackrel{\stackrel{\beta \to 0}{\kappa \to \infty}}{\rightarrow}  \quad 0 \nonumber.
\end{eqnarray}

\section{Conclusions}\label{sec-3}
In this article, it is argued that the dualised field theory for monopoles has a perturbative regime where the coupling of such particles to matter is small due to its dependence on the monopole boost $\beta<<1$ in the centre of mass frame. Monopole models with spins $0$, $\frac{1}{2}$, and $1$ are studied and a new phenomenological parameter $\kappa$ is included, parameterising an unknown magnetic moment contribution to the cross sections. The differential and total cross section distributions are calculated analytically and displayed graphically for monopole pair production by Drell--Yan and photon fusion. In order to make experimentally relevant distributions, these parton-level amplitudes should be convoluted with appropriate parton density functions. Models with different $\kappa$ values are contrasted and the SM-like couplings, $\kappa^{1(1/2)}=1(0)$, which give the bosonic and fermionic monopole moments identical to the SM $W$ boson and electron, respectively, stand out as unitary preserving. PF is clearly seen to dominate over DY at energy scales relevant to current colliders. Finally, allowing $\kappa$ to become large while remaining in the perturbative regime generates production rates by PF accessible to current collider experiments, such as MoEDAL at CERN, in the non-zero spin models.

\begin{acknowledgements}
This research was funded by a Science and Technology Facilities Council (STFC, UK) doctoral studentship. The author would like to thank N. E. Mavromatos and J. R. Ellis for their support in supervising this work. The author would also like to thank V. A. Mitsou, J. Pinfold, and A. Santra as collaborators in \cite{fullreport}, which this report was partly based upon. Last, but not least, the author thanks the MoEDAL Collaboration at CERN for their interest in this project.
The author declares no conflict of interest. The founding sponsors had no role in the design of the study; in the collection, analyses, or interpretation of data; in the writing of the manuscript, or in the decision to publish the results.
\end{acknowledgements}
\vspace{6pt}

\paragraph*{\bf The following abbreviations are used in this manuscript:\newline}

\noindent 
\begin{tabular}{@{}ll}
PF & Photon fusion\\
DY & Drell--Yan\\
SM & Standard Model\\
\end{tabular}

\end{document}